\documentstyle[12pt,epsfig]{article}

\newcommand{\la}{\bar{\Lambda}}
\newcommand{\de}{\Delta}
\newcommand{\m}{\mu_G^2}

\newcommand{\ds}{\displaystyle}
\newcommand{\bq}{\begin {eqnarray} & \ds}
\newcommand{\tr}{& \nonumber \\ &  \ds}
\newcommand{\eq}{& \end {eqnarray}}
\newcommand{\se}{\simeq}
\newcommand{\B}{{\overline M}_B}
\newcommand{\D}{{\overline M}_D}
\newcommand{\bi}{\bibitem}

\begin{document}

\thispagestyle{empty}
\begin{flushright}
{Budker INP-96-23\\
hep-ph/9604376\\
April 1996}
\end{flushright}
\vskip 2cm

\centerline{\bf WHAT ARE THE RIGHT VALUES OF ${\bf{\bar \Lambda}}$}
\vskip 0.5cm
\centerline{\bf AND THE HEAVY QUARK KINETIC ENERGY\,?}
\vskip 1cm
\centerline{\bf Victor Chernyak}
\centerline{ Budker Institute of Nuclear Physics, and Novosibirsk State 
University} 
\centerline{630090\,\,\, Novosibirsk-90, \,\, Russia}
\vskip 1cm

\centerline {\bf Abstract}

The values of two important parameters of the heavy quark effective 
theory,\, $\la$ and $\mu_\pi^2$ (the mean value of the heavy quark three 
momentum squared),\, have been determined recently in \cite{GKLW} from the
precise CLEO data on the shape of the electron spectrum in semileptonic
B meson decays. The values obtained in \cite{GKLW}: $\la=(0.55\pm 0.05)
\,GeV,\, \mu_\pi^2=(0.35\pm 0.05)\,GeV^2,$ 
disagree with the result obtained earlier in \cite{CH} from the D meson
semileptonic width.

The purpose of this note is to show that the main reason for a disagreement
is the secondary electron background present in the data,\, which
influences strongly the extracted values of $\la$ and $\mu_\pi^2$. We
determine the amount of this background from the selfconsistency
conditions,\, and subtract it out. As a result,\, the values of $\la$ and
$\mu_\pi^2$ become a factor two smaller and agree with \cite{CH}.

\newpage
\hskip 1cm {\bf 1}. \\
The value of the charm quark (pole) mass,\, $M_c$,\, has been found in
\cite{CH} from a calculation of the D meson semileptonic width. This is
a good place for finding out a precise value of $M_c$,\, as the KM-factors
$V_{cs},\,V_{cd}$ are
well known,\, the s-quark mass is reasonably well known and,\, in any case,
\, plays a small role here,\, while the Born
contribution behaves as $\sim M_c^5$. Moreover,\, all the main perturbative
and power corrections to the Born term are also sensitive to a precise value 
of $M_c$ and enter the answer with the same (negative) sign,\, which prevents
accidental cancelations.  
So, the decay 
width depends on $M_c$ as: $\Gamma\sim M_c^{n_{eff}}$,\, and $n_{eff}$ is 
noticeably larger than 5. 

The value:
\bq M_c=\bigl [\,1650\pm (\,35\,)_{theor}\pm (\,15\,)_{exp}\bigr ]\,MeV \eq
has been obtained in \cite{CH}. Being combined with the mass formula of the 
heavy quark effective theory ($\,\D=(\,3\,M_{D^*}+M_D\,)/4\,)$:
\bq \D-M_c= \la+\frac{\mu_\pi^2}{2\,M_c}\,,\eq
this gives a tight constraint on the combination of $\la$ and $\mu_\pi^2$ 
(the mean value of the c-quark three momentum squared) entering
the right hand side of Eq.(2):
\bq \la+\frac{\mu_\pi^2}{2\,\D}\left (1+\frac{\la}{\D}\right )=\bigl 
[\,323\pm (\,35\,)_{theor}\pm (\,15\,)_{exp}\bigr ]\,MeV\,. \eq

The first serious attempt has been undertaken recently in \cite{GKLW} to
extract the values of $\bar \Lambda$ and $\mu_\pi^2$ from an independent
sourse: using the precise CLEO data on the shape of the lepton spectrum in 
inclusive semileptonic $B\rightarrow X\,l\,{\bar \nu}$ decays. 
 Much larger values:
\bq \bar \Lambda=0.55\pm 0.05\,GeV\,,\quad \mu_\pi^2=0.35\pm 0.05\,GeV^2\,,\eq
has been obtained in \cite{GKLW},\, in disagreement with Eq.(3).

The purpose of this note is to elucidate the reasons for a discrepancy and 
to present the results of a more careful treatment of CLEO data along the 
lines used in \cite{GKLW}. As a result,\, our values of $\la$ and $\mu_
\pi^2$ extracted from the same data are a factor two smaller than in Eq.(4).
The main reason for such a large difference originates from 
neglecting in \cite{GKLW} the secondary electron background present in the
experimental data. Indeed, for the lepton energy interval used (see below)
this background is small,\, about $1.5\%$. The matter is,\, however,\, that
the parameters ${\bar \Lambda}$ and $\mu_\pi^2$ we are looking for enter the
data as power corrections,\, and their effect is also a few per cent only.
Therefore, the presence of the secondary electron background in the data
used influences strongly the extracted values of ${\bar \Lambda}$ and
$\mu_\pi^2$. Besides,\, we account explicitely for the higher loops 
perturbative corrections (and this also decreases somewhat the value of 
${\bar \Lambda}$),\, and consider in more detail the role of
 third order corrections.  
\vskip 0.5cm
\hskip 1cm {\bf 2.}

Because the results presented in \cite{GKLW} are used heavily below,\, let
us recall in short the line of approach and main definitions. The ratios
are considered:
\bq R_1=\frac{\int_{1.5\,GeV}E_l\,\Gamma(E_l)}{\int_{1.5\,GeV}\Gamma
(E_l)}\,,\quad R_2=\frac{\int_{1.7\,GeV}\Gamma(E_l)}
{\int_{1.5\,GeV}\Gamma(E_l)}\,,\quad R_3=\frac{\int_{1.8\,GeV}\Gamma(E_l)}
{\int_{1.5\,GeV}\Gamma(E_l)}\,,\eq
where $\Gamma(E_l)$ is the differential distribution in the electron energy.
The quantities like $R_{i}$ are most suitable as the largest unknown
factors $M_b^5\,|V_{cb}|^2$ cancel in ratios and,\, besides,\,
the secondary electron background is small at $E_l>1.5\,GeV$,\, while the
role of power corrections we are looking for is enhanced.

The ratios like $R_{i}$ are calculated then theoretically as series in
powers of $\Lambda_{QCD}/M_b$,\, using the operator product expansions and
the heavy quark effective theory. The second order corrections 
to the differential cross section have been
found in \cite{BSUV}\cite{MW},\, while the third order ones have
been calculated recently in \cite{GK}. The results have the form (all 
numbers here and below are given in GeV units):
\bq  R_1^{theor}=1.8061-10^{-2}\Bigl [\delta_1 R_1+\delta_2 R_1+\delta_3 
R_1\Bigr ]\,,\nonumber \eq 
\bq {\hskip -0.5cm} \delta_1 R_1=\Bigl [\,5.82\, \la-8.22\, \mu_\pi^2+
4.67\, \m+1.25\, \la^2- 3.83\, \la\mu_\pi^2-0.24\, \la \m +0.30\, 
\la^3\, \Bigr ]\,,\nonumber \eq 
\bq \delta_2 R_1=\frac{\alpha_s}{\pi}\Bigl (3.5+\frac{7\la}
{\B}\Bigr )\kappa_b^{(w)} -\left 
|10\,\frac{V_{ub}}{V_{cb}} \right |^2\Bigl (1.33-\frac{10.3\la}{\B}\Bigr 
)+\tr +
\Bigl (0.41-\frac{0.4\la}{\B}\Bigr) -\Bigl (0.62+\frac{0.2\la}{\B}\Bigr )\,, 
\nonumber \eq 
\bq \delta_3 R_1=\Bigl [\,5.11\,\rho_1+1.11\,\rho_2+2.15\, 
\de_1-0.05\,\de_2+2.04\,\de_3 \Bigr ]\,. \eq
\bq  R_2^{theor}=0.6584-10^{-2}\Bigl [\delta_1 R_2+\delta_2 R_2+\delta_3 
R_2\Bigr ]\,,\nonumber \eq 
\bq {\hskip -0.5cm} \delta_1 R_2=\Bigl [\,5.92\, \la-5.85\,\mu_\pi^2 
+5.83\, \m+2.40\, \la^2- 4.73\, \la\mu_\pi^2+1.69\, \la \m +1.0\, \la^3\, 
\Bigr ]\,,\nonumber \eq 
\bq  \delta_2 R_2=\frac{\alpha_s}{\pi}\Bigl 
(3.9+\frac{18\la}{\B}\Bigr )\kappa_b^{(w)} -\left 
|10\,\frac{V_{ub}}{V_{cb}}\right |^2\Bigl (0.87-\frac{3.8\la}{\B}\Bigr )+\tr
+\Bigl (0.73+\frac{0.5\la}{\B}\Bigr) -\Bigl (0.21+ \frac{0.3\la}{\B}\Bigr 
)\,, \nonumber \eq 
\bq \delta_3 R_2=\Bigl [\,1.25\,\rho_1+0.59\,\rho_2+1.89\, 
\de_1+0.39\,\de_2+2.66\,\de_3 \Bigr ]\,. \eq
\bq R_3^{theor}=0.4878-10^{-2}\Bigl [\delta_1 R_3+\delta_2 R_3+\delta_3 
R_3\Bigr ]\,,\nonumber \eq 
\bq {\hskip -0.5cm}\delta_1 R_3=\Bigl [\,8.85\, \la-8.78\,\mu_\pi^2 +8.10\, 
\m+3.57\, \la^2- 7.12\, \la\mu_\pi^2+2.26\, \la \m +1.49\, \la^3\, \Bigr 
]\,,\nonumber \eq 
\bq \delta_2 R_3=\frac{\alpha_s}{\pi}\Bigl 
(6.1+\frac{26\la}{\B}\Bigr )\kappa_b^{(w)} -\left 
|10\,\frac{V_{ub}}{V_{cb}}\right |^2\Bigl (\,1.28-\frac{5.7\la}{\B} \,\Bigr 
)+ \tr +\Bigl (\,0.88+\frac{0.4\la}{\B}\,\Bigr) -\Bigl (\,0.39+ 
\frac{0.8\la}{\B}\,\Bigr )\,, \nonumber \eq 
\bq \delta_3 R_3=\Bigl 
[\,1.63\,\rho_1+1.0\,\rho_2+2.73\,\de_1+0.45\,\de_2+3.64\,\de_3 \Bigr ]\,. 
\eq

The terms $\delta_1 R_{i}$ and $\delta_2 R_{i}$ in Eqs.(6-8) have been 
presented in \cite{GKLW}, \footnote{\, The Eq.(8) is not written down 
explicitely in \cite{GKLW} but has been kindly sent on request by Zoltan 
Ligeti to whome I am deeply grateful\,.}
and the terms entering $\delta_3 R_{i}$ are easily 
calculated using the results for the second order terms and those from 
\cite{GK}.  The terms $\delta_1 R_{i}$ represent the first and second order 
corrections,\, and kinematical third order ones. The terms $\delta_2 R_{i}$ 
represent strong and electromagnetic radiative corrections and Lorenz boost 
corrections.  The terms $\delta_3 R_{i}$ are "the dynamical" third order 
corrections.  The nonperturbative parameters entering Eqs.(6-8) are defined 
as follows \cite{M} (the nonrelativistic normalization of states is used,\, 
$\la$ is defined by the matrix element of the light degres of freedom part of 
the Hamiltonian in the infinite mass limit):  
\bq {\hskip -0.7cm}\langle B\,|\,{\bar b}\,{\vec \pi}^2\,b\,|\,B\rangle=\Bigl 
[\,\mu_\pi^2-\frac{(\,\de_1-0.5\,\de_2\,)}{M_b}\,\Bigl ]\,,\nonumber \eq
\bq \langle B\,|\,{\bar b}\,\bigl({\vec \sigma}\,g_s{\vec H}\,\bigl 
)\,b\,|\,B\rangle=\Bigl[\,-\m-\frac{(\,\de_3-\de_4-0.5\,\de_2\,)}{M_b}\,
\Bigl ]\,, \nonumber  \eq 
\bq \langle B\,|\,{\bar b}\,\pi_{\alpha}\,\pi_{\mu}\,\pi_{\beta}\,b\,|\,B 
\rangle=\frac{\rho_1}{3}\,v_{\mu}\,\bigl (\,g_{\alpha\beta}-v_{\alpha} 
v_{\beta}\,\bigl )\,, \nonumber \eq
\bq \langle B\,|\,{\bar 
b}\,\pi_{\alpha}\,\pi_{\mu}\,\pi_{\beta}\,\gamma_{\delta}\,\gamma_5\,b\,|\,B 
\rangle=\frac{i\rho_2}{6}\,v_{\mu}\,
\epsilon_{\alpha\beta\nu\delta}\,v_\nu\,.\eq

Here: $ \pi_\mu$ is the heavy quark momentum operator,\,
the terms $\de_i$ originate from the corrections to
the B meson wave function and are naturally expressed through the 
corresponding two point correlators,\, while the terms $\rho_{1,2}$ are the 
genuin local third order corrections. Being expressed in more visible terms 
they look as:  
\bq \rho_1=-\frac{1}{2}\langle B\,|\,{\bar b}\,\bigl(\,g_s{\vec 
D}{\vec E}\,\bigr )\,b\,|\,B \rangle \se \tr \se -2\,\pi\,\alpha_s \langle 
B\,|\,{\bar b} \,\frac{\lambda^a}{2}\,\gamma_\mu\,b\cdot {\bar 
q}\,\frac{\lambda^a}{2}\,\gamma_\mu \,q\,|\,B\rangle \se 
\frac{2}{9}\pi\alpha_s M_B f^2_B\,,\eq 
\bq \rho_2=\langle B\,|\,{\bar b}\,{\vec \sigma}\bigl (\,g_s{\vec E}\,
\times {\vec \pi}\bigl )\,b\,|\,B\rangle\,.\eq

In terms of the above parameters the meson masses look as (\,
$\B=(3\,M_{B^*}+M_B)/4$,\,): 
\bq \B=M_b+\la+\frac{\mu_\pi^2}{2\,M_b}+\frac{(
\rho_1-\de_1-\de_3)}{4\,M_b^2}\,,\eq
\bq \frac{3}{2}\Bigl (M_{B^*}-M_B\Bigl )\B=\Bigl [\,\m-\frac{\Delta_0}
{2\,M_b}\,\Bigl ]\,,\eq
\bq \frac{3}{2}\Bigl (M_{D^*}-M_D\Bigl )\D\,\eta=
\Bigl [\,\m-\frac{\Delta_0}{2\,M_c}\Bigl ]\,\eq
\bq \Delta_0=\bigl (\,\de_2+\de_4+\rho_2-2\,\m \la\,\bigl )\,,\,\,
 \eta=\left (\frac{\alpha_s(M_b)}{\alpha_s(M_c)}\right )^{\frac{9}{25}}
\se 0.86\,.\eq

As for other parameters entering Eqs.(6-8),\, we use: $\alpha_s=\alpha_s(M_b)
=0.21,\, |10\,V_{ub}/V_{cb}|=0.8,\,$ while the parameter $\kappa_b^{(w)}$
describes the summary effect of the Borel ressumed perturbation theory
corrections. Its characteristic value for the B meson semileptonic decay is
\cite{BB}: $\kappa_b^{(w)}=2.1\,.$ Besides,\, as the left hand sides of
Eqs.(13,\,14) are known,\, we have:
\bq \m\se 0.36\,,\quad \Delta_0\se 0\,,\quad \bigl (\,\de_2+\de_4
+\rho_2\,)\se 2\,\m \la\,.\eq
Finally,\, we use below the value: $\rho_1\se 0.012,\,$ which corresponds
(\,see Eq.(10)\,) to $f_B\se 0.12$ found in \cite{CH}.
\vskip 0.5cm
\hskip 1cm {\bf 3}.

The experimental values of the ratios $R_i$ in Eqs.(6-8) are 
\footnote{\, The systematic errors are not considered
and are expected to cancel to a large extent in the ratios Eq.(5),\, see 
\cite{GKLW},\, the statistical errors will be accounted for below\,.}:  
\bq R_1^{exp}=1.7830=R_1^{theor}(1-\sigma_1)\,,\quad 
R_2^{exp}=0.6108=R_2^{theor}(1-\sigma_2)\,,\tr
R_3^{exp}=0.4276=R_3^{theor}(1-\sigma_3)\,,\eq 
where $\sigma_i$ denote 
possible background contributions of secondary electrons. These can be found 
as follows. Equating the expressions of $R_2^{theor}$ and 
$R_3^{theor}$ from Eqs.(7,8) and Eq.(17),\, these can be rewritten as:  
\bq \Bigl [\,\la-0.85\,\mu_\pi^2+0.84\,\m+0.35\,\la^2-0.69\,
\mu_\pi^2\la+0.24\,\m\la +0.15\,\la^3+ 
\tr +0.18\,\rho_1+0.09\,\rho_2+0.27\,\de_1+0.05\,\de_2+0.39\,\de_3\,\Bigl ]= 
\tr =\Bigl [\,0.616-8.86\,\sigma_2\,\Bigl ]\,,\eq 
\bq \Bigl [\,\la-0.87\,\mu_\pi^2+0.80\,
\m+0.35\,\la^2-0.70\,\mu_\pi^2\la+0.22\,\m\la +0.15\,\la^3+ 
\tr +0.16\,\rho_1+0.10\,\rho_2+0.27\,\de_1+0.04\,\de_2+0.36\,\de_3\,\Bigl ]= 
\tr =\Bigl [\,0.541-4.21\,\sigma_3\,\Bigl ]\,,\eq

It is seen that the left hand sides are (nearly) equal. So,\, the right
hand ones should be equal as well. Besides,\, it is clear that the
secondary electron background is really small for $E_l>1.7\,GeV$,\, and
originates mainly from the interval $1.5\,GeV<E_l<1.7\,GeV$. So,\,
$\sigma_2$ and $\sigma_3$ should be close to each other. Taking $\sigma_2
\se \sigma_3$,\, we obtain from Eqs.(18,19) as a first approximation: 
$\sigma_2\se \sigma_3\se 1.6\%$.
We need also $\sigma_1$,\, which can be found now as follows.

Because we deal with the very end of the secondary electron spectrum,
\, its form can be well approximated by a simplest stright line:
\bq \frac{1}{\Gamma_0}\,\delta 
\Gamma(E_l)=C_0\,(\,1.78-E_l\,)\,\theta(1.78-E_l), \quad  
\Gamma_0=\int_{1.5\,GeV} \Gamma^{theor}(E_l)\,,\eq
where $\delta \Gamma(E_l)$ is the contribution to the differential cross
section from secondary electrons. It is not difficult to obtain then:
\bq \sigma_3=3.9\%\cdot C_0\,,\quad \sigma_2=3.4\%\cdot C_0\,,\quad
\sigma_1=0.45\%\cdot C_0\,.\eq
Choosing now the coefficient $C_0=0.385$ from (see above) $\sigma_3=1.5\%$,\,
we obtain: 
\bq \sigma_1=0.17\%\,, \quad \sigma_2=1.3\%\,, \quad \sigma_3=1.5\%\,.\eq
As a check of the above values of the secondary electron background,\, we can
estimate also the amount of this background for the $E_l>1.4\,GeV$ 
electrons and obtain $\se 2.4-2.5\%$,\, which compares well with the 
CLEO value $(\,2.8\pm 0.7\,)\%$ \cite{CLEO}.

Let us emphasize,\, that the above found values of the secondary electron
background are model independent as they are obtained solely from the
selfconsistency requirements of the above written equations,\, i.e. 
requiring that $R_2$ and $R_3$ give the same result.
\footnote{
We see no reasons to question the validity of the quark-hadron duality for
the $E_l>1.8\,GeV$ electrons and to trust it simultaneously for the
$E_l>1.7\,GeV$ ones. Let us recall,\, that even with $E_l>1.8\,GeV$ we are
summing over the hadron masses from $M_D$ and up to $\se 3.15\,GeV,\,$ 
covering thus a sufficiently large number of states. } Let us repeat 
also that the background subtraction influences strongly the 
extracted values of $\la$ and $\mu_\pi^2$. First,\, it is seen from 
Eqs.(18,19,22) that the background is not very small really by itself. Its 
role is strenthened additionally by the fact that the curves obtained from 
$R_1$ and $R_2$ (or from $R_1$ and $R_3$ which are the same now) intersect 
at a small angle,\, and so the position of the intersection point is 
sensitive to such corrections.  \vskip 0.5cm \hskip 1cm {\bf 4}.

Let us proceed now to some numerical results which follow from the above 
equations. As a zeroth approximation,\, we can neglect all third order 
corrections,\, both kinematical and dynamical ones,\, and obtain then from
Eqs.(6,7,17,22) for the central values
\footnote{\, As $R_2$ and $R_3$ 
give identical results after the background is subtracted out,\, 
 we deal with the $(\,R_1,\, R_2\,)$ pair for a definitness\,.}: 
\bq \la^{(0)}=0.310\,, \quad \mu^{2,(0)}_\pi=0.175\,. \eq
It is seen that the results for both $\la$ and $\mu_\pi^2$ are a factor two
smaller in comparison with Eq.(4),\, and this is mainly due to a background
subtraction.

Let us consider now in some detail a possible role of the third order terms.
As was noticed above,\, the terms $\delta_1 R_{1,2}$ contain kinematical 
corrections: $\m\la$,\, $\mu_\pi^2\la$ and $\la^3$,\, and nothing 
prevents us from accounting for these. 
Accounting also for $\rho_1\se 0.012$,\, one obtains now:
\bq \la=0.280\,,\quad \mu_\pi^2=0.135\,, \eq
($\,\la=0.265,\, \mu_\pi^2=0.115\,$ with $\rho_1=0\,$), which can
be compared with the values: $\la=0.500\,,\,\, \mu^2_\pi=0.270\,,$
obtained in \cite{GKLW} in a similar approximation. 

The dynamical third order terms $\de_i$ are unknown,\, of course.
A hint on their possible values is given,\, however,\, by Eq.(16) which
shows that,\, with the above used definitions,\, they are naturally positive
and of a natural size: $\sim \m\la\se 0.1$,\, as one could
expect beforehand. \footnote{\, Let us recall \cite{CH} that,\, 
analogously to the $<(\,{\vec \sigma}\,{\vec \pi}\,)^2>$ matrix element 
and unlike the quantum mechanics,\, there are no positiveness conditions
for "the genuine nonperturbative terms" $\de_i$,\, in spite of that some
bilocal correlators look positive definite. As usual,\, there are power 
divergent loop corrections in these correlators which should be subtracted 
out,\, and the terms $\de_i$ represent "what is left". 
Clearly,\, "what is left" depends essentially on the subtraction scheme. 
We don't share the optimistic viewpoint that,\, i.e. with the upper cut off 
$\mu\se 1\,GeV,$\, "what is left" is much larger than the subtracted part. 
This later,\, for instance,\, contributes typically $\sim 
(\alpha_s(\mu)\,\mu^3/\pi)\se 0.15$ to the correlator $<{\vec \pi}^2,\,
{\vec \pi}^2>,\,$ that is of the same size as $\m\la$,\, see Eq.(16)\,.}

To illustrate their possible role we give below a number
of examples (compare with Eq.(24)),\, 
taking various natural values for these parameters 
($\rho_1$ is always fixed at 0.012):  
\bq 1)\,\,\,\de_1=0.1\,,\,\, \de_2=\de_3=\rho_2=0\,:
\quad\la=0.270\,,\quad \mu_\pi^2=0.150\,, \nonumber \eq 
\bq 2)\,\,\,\de_3=0.1\,,\,\, \de_2=\de_3=\rho_2=0\,: 
\quad \la=0.240\,,\quad \mu_\pi^2=0.125\,, \nonumber \eq
\bq 3)\,\,\,\de_2=0.2\,,\,\, \de_1=\de_3=\rho_2=0\,: 
\quad \la=0.250\,,\quad \mu_\pi^2=0.110\,,\nonumber \eq
\bq 4)\,\,\,\rho_2=0.2\,, \,\,\de_1=\de_2=\de_3=0\,: 
\quad \la=0.295\,,\quad \mu_\pi^2=0.170\,, \nonumber \eq
\bq 5) \de_2=\rho_2=0.1\,,\,\, \de_1=\de_3=0\,:
\quad \la=0.270\,,\quad \mu_\pi^2=0.140\,. \eq

It is seen that,\, in comparison with Eq.(24),\, $\de_{1,2,3},\,\rho_2
\neq 0$ give reasonably small corrections.

Moreover,\, it is commonly believed that because $\rho_2$ 
originates from the spin-orbital interaction,\, see Eq.(11),\, its  real
value is suppressed for the ground state B meson,\, so that it is more
realistic that it is of the same order as $\rho_1,\,$ 
rather than $\sim 0.1-0.2$. 

Because $\m$ is considerably larger 
than $\mu_\pi^2,\,$ one can expect also that 
the terms $\de_3$ and $\de_4$ are potentially the largest ones. But $\de_3$ 
only decreases the answer (see Eq.(25)),\, while $\de_4$ is already accounted 
for in Eqs.(6-8),\, as it is substituted by $\de_4= (2\m\la-\de_2-\rho_2)$ 
from Eq.(16),\, so that Eq.(24) corresponds really to the preferable case: 
$\de_4\se 0.2,\, \de_1\se \de_2\se \de_3\se \rho_2\se 0.$

In any case,\, varying third order terms within their natural limits we can
see that their effect is small and is typically within the experimental
statistical error bars (see below).

The effect due $\kappa_b^{(w)}\neq 1$ in
Eqs.(6-8) is also very mild. For instance,\, one obtains (with
$\de_{1,2,3}=\rho_2=0,\, \rho_1=0.012$,\, compare with Eq.(24)): 
$\la=0.330,\, \mu_\pi^2=0.145,\,$ even with the unrealistic value: 
$\kappa_b^{(w)}=1$. This shows that the value 
of $\kappa_b^{(w)}$ influences mainly $\la$,\, while $\mu_\pi^2$ stays 
nearly intact. 
This is as expected,\, as varying $\kappa_b^{(w)}$ is equivalent,\, in 
essence,\, to the renormalon redefinition,\, i.e. changing the summation 
prescription for the divergent perturbation theory. And it is $\la$ which is 
affected by the leading renormalon,\, while $\mu_\pi^2$ is not.  
\footnote{\, 
Really,\, one can expect the precise value of $\kappa_b^{(w)}$ to be even a 
bit larger than the value 2.1 which we use and which corresponds to the total 
width. Indeed,\, as we deal with $E_l>1.5\,GeV$,\, the radiated gluons are 
softer and this will lead to a larger value of $\kappa_b^{(w)}$,\, in 
comparison with the total width.} 

Finally,\, varying $|V_{ub}/V_{cb}|$ also gives a small effect. One obtains:
$\la=0.290,\, \mu_\pi^2=0.115$ instead of Eq.(24) at $|V_{ub}/V_{cb}|=0.1.$
\vskip 0.5cm
\hskip 1cm {\bf 5}.

The statistical errors of CLEO data are given by the correlation 
matrix \cite{GKLW}:  
\bq V(\,R_1,\,R_2\,)=\left ( \matrix { 
1.64\times 10^{-6} & 2.08\times 10^{-6} \cr 
2.08\times 10^{-6} & 5.45\times 10^{-6}
}\right ) \eq
Taking the case of Eq.(24) as a central point,\, one obtains the
figure which is quite similar to Fig.1 in \cite{GKLW},\, but with the
central point at $\la=280\,MeV,\,\,\mu_\pi^2=0.135\,GeV^2$. Eq.(3),\, which
is obtained from the D meson semileptonic decays and is completely 
independent,\, gives the additional band. We don't even try here to
write out "the right central values" of $\la$ and $\mu_\pi^2$ and,\,
especially,\, "the right error bars" which follow from all the above
described results. Rather,\, this is a problem for a specialist. 
 As a typical example,\, we show only in Fig.1 the central lines of the 
case of Eq.(24),\, together with the central line of Eq.(3).  (Two nearly 
coinciding lines are from $R_2$ and $R_3$,\, those which intersects them 
at a small angle is from $R_1$,\, and those at a large one is from Eq.(3)).

Nevertheless,\, clearly,\, the final results of the form:
\bq \la=\bigl (\,280\pm 40\,\bigr )\,MeV\,,\quad 
\mu_\pi^2=\bigl (\,0.14\pm 0.03\,)\,GeV^2\,,\eq
are definitely close to the "right" ones. 

Most surprising is the small value of $\mu_\pi^2$ which is more than
three times smaller the widely used value $\mu_\pi^2\se 0.5\,GeV^2$,\,
and is small even in comparison with the value $0.25\,GeV^2$ used in
\cite{CH}. As a
result,\, the difference of the quark (pole) masses looks now as:
\bq M_b-M_c=\bigl (\,3370\pm 10\,\bigr )\,MeV\,,\eq
while the central values of $M_c$ and $M_b$ which follow from Eq.(27) are:
\bq M_c=1650\,MeV\,,\quad M_b=5020\,MeV\,,\eq
and agree with those obtained in \cite{CH}. As for $|V_{cb}|$,\, proceeding
in the same way as in \cite{CH},\, one obtains the result:
\bq |V_{cb}|\cdot 10^3=(\,42.5\pm 1\,)\left [\frac{Br(B_d\to l\nu+X)}
{11.0\%}\right ]^{1/2}\left [\frac{1.6\,ps}{\tau(B_d)}\right ]^{1/2}\eq
which coincides practically with those obtained in \cite{CH},\, 
and only receives now more confidence.

Let us comment finally in short on a comparison of the above result,\, Eq.
(27),\, with those obtained in \cite{FLS} 
from a calculation of the hadron invariant mass distributions
in the B meson semileptonic decays. At present,\, the weak point of this
approach is a poor accuracy of experimental data on a production of
$D^{**}$ states in B decays. The result: $\la\se 450\,MeV$ \cite{FLS}
relies heavily on the OPAL data which gave highest production rates of the
$D^{**}$ states. At the same time,\, ALEPH,\, DELPHI and CLEO all indicate
smaller production rates. It is not difficult to check that it is
sufficient to diminish the OPAL central values on $\sim 2\sigma$ to avoide
disagreement with the above results,\, Eq.(27). Clearly,\, as the quality
of the experimental data will improve,\, the results obtained within the 
approach used in \cite{FLS} will become more reliable. 

Some caution is needed,\, however,\, 
when comparing our results with those from
\cite{FLS}. These authors restrict themselves to two loop perturbative
corrections. This corresponds to smaller value of $\kappa_b^{(w)}$,\, in
comparison with those we use and which corresponds to a Borel resummed
perturbation series. Being considered as a redefinition of the summation
prescription for a divergent perturbative series,\, this will correspond
to a redefinition of $\la$,\, so that their $\la$ is a bit larger in
comparison with our one.

\newpage
\begin{figure}
\mbox{\epsfig{file=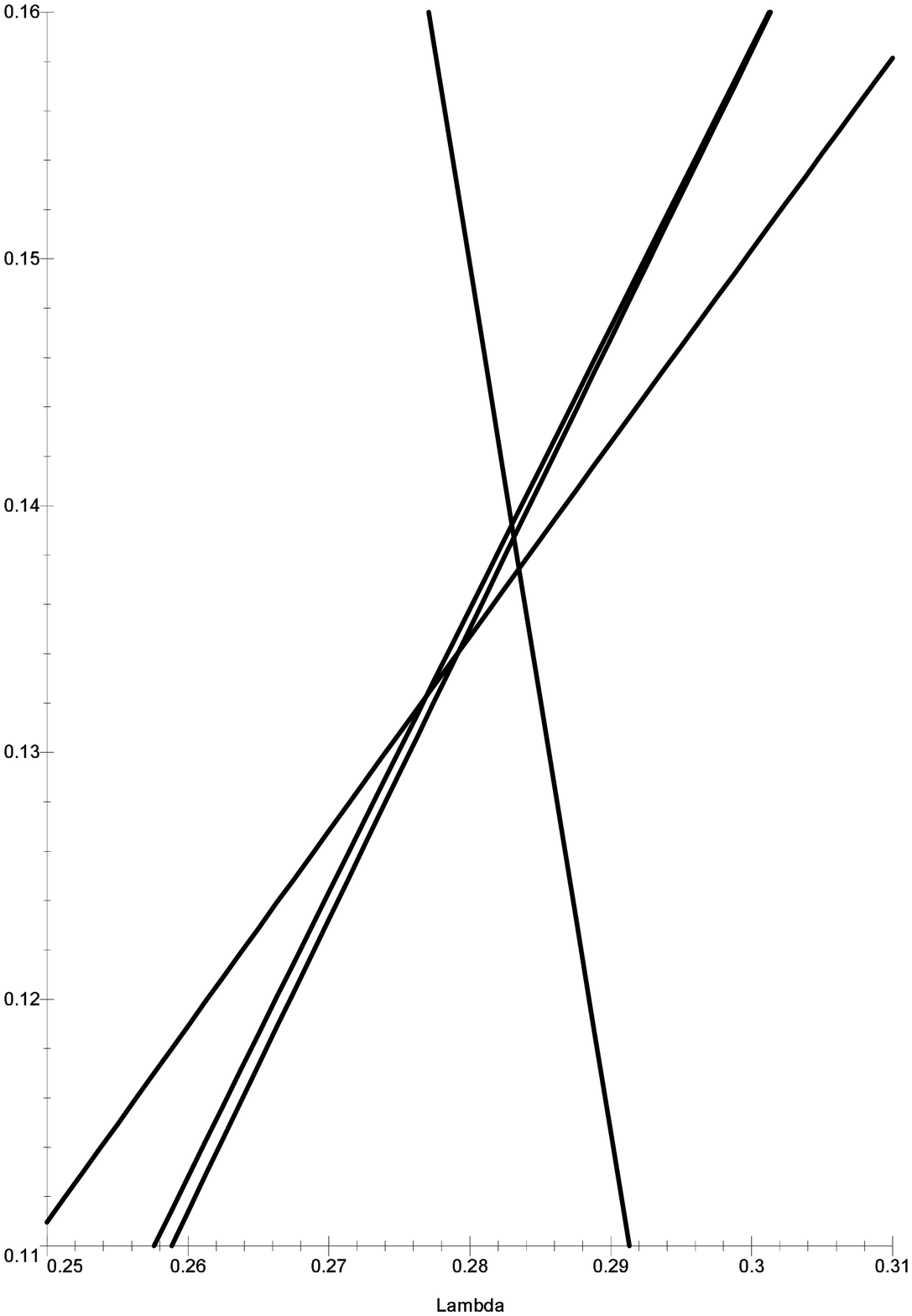,width=\textwidth}}

\end{figure}

\end{document}